\newif\ifpreprint

\preprintfalse 

\ifpreprint
\documentclass[journal=jctcce,manuscript=letter]{achemso}
\else
\documentclass[journal=jctcce,manuscript=letter,layout=twocolumn]{achemso}
\fi

\usepackage[T1]{fontenc} 

\usepackage{amsmath}
\usepackage{newtxtext,newtxmath}

\usepackage{graphicx}
\usepackage{dcolumn}
\usepackage{braket}
\usepackage{multirow}
\usepackage{threeparttable}
\usepackage{xspace}
\usepackage{verbatim}
\usepackage[version=4]{mhchem} 
\usepackage{comment}
\usepackage{color,soul}
\usepackage{siunitx}
\usepackage{physics}

\usepackage{mathtools}
\usepackage[dvipsnames]{xcolor}
\usepackage{xspace}
\usepackage{ifthen}

\usepackage{qcircuit}

\usepackage{graphicx,longtable,dcolumn,mhchem}
\usepackage{rotating,color}
\usepackage{lscape}
\usepackage{amsmath}
\usepackage{dsfont}
\usepackage{soul}
\usepackage{physics}
\newcolumntype{d}{D{.}{.}{-1}}

\usepackage[normalem]{ulem}

\usepackage[colorlinks = true,
            linkcolor = blue,
            urlcolor  = blue,
            citecolor = blue,
            anchorcolor = blue]{hyperref}
\definecolor{goodorange}{RGB}{225,125,0}
\definecolor{goodgreen}{RGB}{5,130,5}
\definecolor{goodred}{RGB}{220,50,25}
\definecolor{goodblue}{RGB}{30,144,255}

\newcommand{\note}[2]{
\ifthenelse{\equal{#1}{F}}{
\colorbox{goodorange}{\textcolor{white}{\footnotesize \fontfamily{phv}\selectfont #1}}
    \textcolor{goodorange}{{\footnotesize \fontfamily{phv}\selectfont #2}}\xspace
}{}
\ifthenelse{\equal{#1}{R}}{
\colorbox{goodred}{\textcolor{white}{\footnotesize \fontfamily{phv}\selectfont #1}}
    \textcolor{goodred}{{\footnotesize \fontfamily{phv}\selectfont #2}}\xspace
}{}
\ifthenelse{\equal{#1}{N}}{
\colorbox{goodgreen}{\textcolor{white}{\footnotesize \fontfamily{phv}\selectfont #1}}
    \textcolor{goodgreen}{{\footnotesize \fontfamily{phv}\selectfont #2}}\xspace
}{}
\ifthenelse{\equal{#1}{M}}{
\colorbox{goodblue}{\textcolor{white}{\footnotesize \fontfamily{phv}\selectfont #1}}
    \textcolor{goodblue}{{\footnotesize \fontfamily{phv}\selectfont #2}}\xspace
}{}
}

\usepackage{titlesec}

\usepackage[fontsize=11pt]{scrextend}
\titleformat{\section}
{\normalfont\sffamily\bfseries\color{Blue}}
{\thesection.}{0.25em}{\uppercase}

\titleformat{\subsection}
{\normalfont\sffamily\bfseries}
{\thesubsection}{0.25em}{}

\titleformat{\subsubsection}
{\normalfont\sffamily}
{\thesubsubsection}{0.25em}{}

\titleformat{\suppinfo}
{\normalfont\sffamily\bfseries}
{\thesubsection}{0.25em}{}

\titlespacing*{\section}{0pt}{0.5\baselineskip}{0.01\baselineskip}
\titlespacing*{\subsection}{0pt}{0.125\baselineskip}{0.01\baselineskip}
\titlespacing*{\subsubsection}{0pt}{0.125\baselineskip}{0.01\baselineskip}

\newcommand{\CEISAM}{Nantes Universit\'e, CNRS,  CEISAM UMR 6230, F-44000 Nantes, France}
\newcommand{\ILM}{Université de Lyon, Université Claude Bernard Lyon 1, CNRS, Institut Lumi\`ere Mati\`ere, F-69622, Villeurbanne, France}

\author{Lina Fransén}
	\affiliation[UN, Nantes]{\CEISAM}
\author{Thierry Tran}
	\affiliation[UN, Nantes]{\CEISAM}
\author{Saikat Nandi}
	\affiliation[ILM, Lyon]{\ILM}    
\author{Morgane Vacher}
	\email{morgane.vacher@univ-nantes.fr}
	\affiliation[UN, Nantes]{\CEISAM}    
		
\let\oldmaketitle\maketitle
\let\maketitle\relax
     \title{Dissociation and isomerization following ionization of ethylene: insights from non-adiabatic dynamics simulations}
\date{\today}

\begin{tocentry}
	\centering
	\includegraphics[width=.65\textwidth]{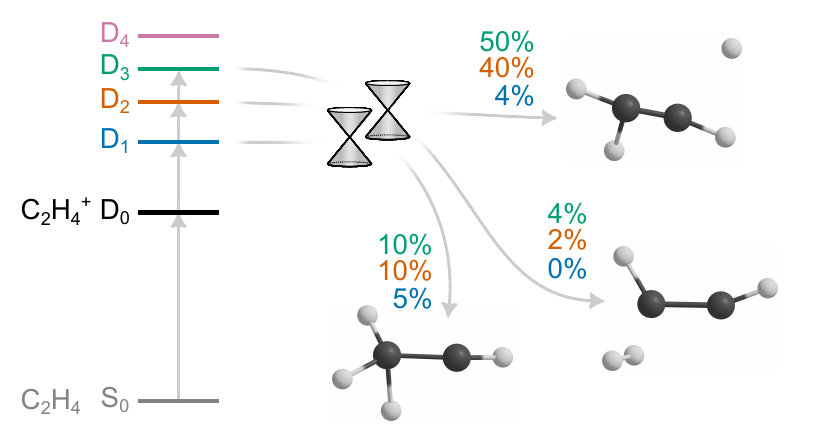}
\end{tocentry}

\begin{document}

\ifpreprint
\else
\twocolumn[
\begin{@twocolumnfalse}
\fi
\oldmaketitle

\begin{abstract}
Photoionized and electronically excited ethylene \ce{C2H4+} can undergo \ce{H}-loss, \ce{H2}-loss, and ethylene-ethylidene isomerization, where the latter entails a hydrogen migration. Recent pioneering experiments with few-femtosecond extreme ultraviolet pulses and complementary theoretical studies have shed light on the photodynamics of this prototypical organic cation. However, no theoretical investigation based on dynamics simulations reported to date has described the mechanisms and time scales of dissociation and isomerization. Herein, we simulate the coupled electron-nuclear dynamics of ethylene following vertical ionization and electronic excitation to its four lowest-lying cationic states. The electronic structure is treated at the CASSCF level, with an active space large enough to describe bond breaking and formation. The simulations indicate that dissociation and isomerization take place mainly on the cationic ground state and allow the probing of previous hypotheses concerning the correlation between the photochemical outcome and the traversed conical intersections. The results, moreover, support the long-standing view that \ce{H2}-loss may occur from the ethylidene form. However, the ethylene-ethylidene isomerization time predicted by the simulations is considerably longer than those previously inferred from indirect experimental measurements.
\end{abstract}

\ifpreprint
\else
\end{@twocolumnfalse}
]
\fi

\ifpreprint
\else
\small
\fi

\noindent

\section{Introduction}
\label{sec:Intro}

The ethylene cation, the simplest organic $\pi$ radical, has served as a prototype system in experimental and theoretical investigations on the mechanism of electronic relaxation, structural dynamics, vibronic coupling, and energy redistribution\cite{Stockbauer-1975, Pollard-1984, Willitsch-2004, Lorquet-1980, Sannen-1981, Kim-2005, Kim2007, Joalland-2014, Tilborg-2009, Ludwig-2016, Zinchenko-2021, Vacher-2022, Lucchini-2022}. Dynamics simulations have predicted that electronically excited \ce{C2H4+} relaxes to the lowest-lying doublet state through conical intersections (CIs) associated with twisted and planar geometries\cite{Joalland-2014, Ludwig-2016, Zinchenko-2021}, and dissociation via \ce{H}- and \ce{H2}-loss has been demonstrated experimentally\cite{Stockbauer-1975}. Elimination of \ce{H2} has, based on static simulations, been suggested to proceed via isomerization to the ethylidene form \ce{CH3CH+}\cite{Sannen-1981}. 

A number of recent experimental works have utilized the time resolution afforded by emerging technologies to investigate the non-adiabatic dynamics of photoionized and electronically excited ethylene\cite{Tilborg-2009, Ludwig-2016, Vacher-2022, Zinchenko-2021, Lucchini-2022}. Zinchenko el al. \cite{Zinchenko-2021} observed by attosecond transient absorption spectroscopy that electronic relaxation between the low-lying states of \ce{C2H4+} takes place in less than 7\,fs\cite{Zinchenko-2021}. Several experimental\cite{Tilborg-2009} and combined experimental-theoretical\cite{Ludwig-2016, Vacher-2022, Lucchini-2022} investigations on this system have moreover employed XUV-pump NIR-probe schemes, where the pump is an attosecond pulse train generated by high-harmonic generation (HHG)\cite{Nisoli-2017}, and the probe is a replica of the HHG driving pulse. The spectrally broad XUV pump ionizes neutral ethylene and launches a nuclear wave packet on several cationic electronic states simultaneously; the probe provides indirect information on the ensuing relaxation dynamics by inducing molecular fragmentation. Challenges associated with this experimental protocol include (i) disentangling the many parallel relaxation pathways and (ii) rationalizing the time-dependent fragment ion yields. Complementary theoretical studies, and in particular non-adiabatic dynamics simulations, have proved useful in the interpretation of the experimental data\cite{Ludwig-2016, Vacher-2022, Lucchini-2022}. Recently, a comparison between an experimental and a simulated isotope effect was shown to aid the identification of the electronic states and nuclear coordinates involved in the relaxation process\cite{Vacher-2022}.

Considerable research effort has been devoted to understanding the mechanisms of ethylene-ethylidene isomerization, \ce{H}-loss, \ce{H2}-loss. Using the XUV-pump NIR-probe scheme described above, Tilborg et al. \cite{Tilborg-2009} inferred experimentally an ethylene-ethylidene isomerization time of 50$\pm$25\,fs\cite{Tilborg-2009}. This isomerization time was later refined, also experimentally, to 30$\pm$3\,fs by Ludwid et al. \cite{Ludwig-2016}, who employed an experimental setup with higher time resolution\cite{Ludwig-2016}. From a theoretical point of view, no statistically significant amount of \ce{H}- or \ce{H2}-loss was predicted by the TD-DFT/PBE0 non-adiabatic dynamics simulations supplementing the experiments in \cite{Ludwig-2016}, nor was ethylene-ethylidene isomerization described theoretically. Similarly, Joalland et al. \cite{Joalland-2014} note in their purely theoretical work, where the electronic structure was treated at the complete active space self-consisistent field (CASSCF) level with active spaces comprising 11 electrons in 7 orbitals and 11 electrons in 8 orbitals, that "No dissociation events have been observed from the excited states, nor any H migrations, although dissociation might be possible with a more flexible wave function (i.e., with a larger active space)"\cite{Joalland-2014}.

Joalland et al. \cite{Joalland-2014}, nevertheless, hypothesized factors that may govern the competition between \ce{H}- and \ce{H2}-loss. They identified two distinct classes of CIs, characterized by planar and twisted geometries. The population transfer through the latter class was shown to be enhanced at high excitation energies. To rationalize an experimentally observed decrease in the \ce{H2}/\ce{H}-loss ratio at high excitation energies\cite{Kim2007}, Joalland et al. \cite{Joalland-2014} therefore proposed that the twisted relaxation channel may suppress ethylene-ethylidene isomerization (and by extension \ce{H2}-loss) by inhibiting the required vibrational energy redistribution; transitions at twisted geometries were associated with an excitation of the vibrational mode $\nu_4$, which, because of symmetry, may not mix with the other modes. A planar CI located along the \ce{H}-migration coordinate was, conversely, proposed to favor ethylene-ethylidene isomerization and \ce{H2}-loss. These hypotheses are in contradiction with the views of \cite{Lorquet-1980} and \cite{Sannen-1981}, who based on static calculations proposed that excitation of $\nu_4$ is required for ethylene-ethylidene isomerization\cite{Lorquet-1980}, and that the competition between \ce{H}- and \ce{H2}-loss is governed by a CI along a \ce{C-H} stretching coordinate\cite{Sannen-1981}.

The details of the dissociation and isomerization dynamics of the ethylene cation and the influence of the CIs on this photoreactivity are still not well understood. In this work, we simulate the dissociation and isomerization dynamics of ethylene following ionization and excitation to its four lowest-energy cationic electronic states using surface hopping, a mixed quantum-classical non-adiabatic dynamics method. The electronic structure is described at the CASSCF level, with an active space comprising all valence electrons (11 electrons in 12 orbitals). The large active space allows a description of bond breaking, and thereby an exploration of the mechanisms, yields, and time scales of \ce{H}-loss, \ce{H2}-loss, and ethylene-ethylidene isomerization. The simulation results moreover enable a revisiting of previous hypotheses concerning the photochemical outcomes favored by relaxations through twisted and planar CIs. 

This article is organized as follows. The theoretical methods are detailed in the next section. The subsequent section presents the results and discusses the findings in the context of previously published experimental and theoretical works. The last section then concludes the article. 

\section{Theoretical methods} 
\label{sec:Methods}
All electronic structure and non-adiabatic dynamics calculations were carried out using the OpenMolcas\cite{Open-Molcas-2019, Open-Molcas-2023} software (version 22.10-354-g7f2c128\cite{v22.10}). 

\subsection{Electronic structure}
The electronic structure of the ethylene cation was treated with the CASSCF method\cite{Roos-1980} with state-averaging over the five lowest-energy cationic states. An active space comprising 11 electrons in 12 orbitals was employed: as shown by some of the present authors, this active space is large enough to describe bond breaking and formation.\cite{Vacher-2022} The following orbitals were included: the $\sigma$ and $\sigma^*$ orbitals of the four \ce{C-H} bonds, and the $\sigma$, $\sigma^*$, $\pi$ and $\pi^*$ orbitals of the \ce{C-C} bond (see Figure S1 in the SI). Only the \ce{C} 1s orbitals were thus inactive. The effect of adding dynamic electron correlation through XMS-CASPT2 was tested with scans along key coordinates (Figures S2-S4 in the SI): in general, no significant differences were observed. The atomic compact Cholesky decomposition\cite{Aquilante-2009} was used to reduce the computational cost, and the basis set of choice was the atomic-natural-orbital relativistic with core correlation basis set with polarized triple-$\zeta$ contraction (ANO-RCC-VTZP)\cite{Roos-2004}. 

Minimum energy conical intersections (MECIs) were optimized at the above-cited level of theory starting from $\sim$150 \ce{D1}/\ce{D0} hopping geometries from the dynamics simulations as well as from previously reported structures\cite{Joalland-2014}. Moreover, the neutral ground state was optimized and associated harmonic frequencies were calculated using an active space of 12 electrons distributed in 12 orbitals. Here, only the lowest-energy root was considered. 

\subsection{Non-adiabatic dynamics}
Initial conditions were generated using Newton-X\cite{Barbatti-2014} by sampling 300 geometries and velocities in an uncorrelated fashion from the Wigner distribution\cite{Wigner-1932} using the frequencies computed at the optimized neutral ground state. For each of the 300 pairs of geometries and velocities, one trajectory was initiated on each of the \ce{D1}, \ce{D2}, and \ce{D3} states. 100 trajectories were initiated on \ce{D0}. This approach treats the electronic superposition as an incoherent one.

Non-adiabatic dynamics simulations were carried out using the surface hopping method employing the Tully fewest switches algorithm\cite{Tully-1990}. Hops between the five lowest-energy cationic states were allowed. The trajectories were propagated with a nuclear time step of 20\, a.u. (0.48\,fs) for $\sim$200\,fs (203\,fs); adequate energy conservation with this time step is demonstrated in Figure S5 in the SI. Each nuclear time step was split into 96 substeps for the propagation of the electronic wave function, which was performed with the Hammes-Schiffer Tully scheme\cite{Hammes-Schiffer-1994} using biorthonormalization, as recently implemented in OpenMolcas\cite{Merritt-2023}. The energy-based decoherence correction devised by Persico and Granucci\cite{Granucci-2007} was employed, with a decay factor of 0.1. 

\subsection{Analysis}
A small number of trajectories, amounting to 0.0, 0.3, 3.3, and 7.3\% of those initiated on \ce{D0}, \ce{D1}, \ce{D2} and \ce{D3}, respectively, exhibit a change in total energy of more than 0.5\,eV during the dynamics. These trajectories were discarded, and the analyses include the remaining 967 trajectories. A statistical ensemble is essential for exploring the relationship between the photochemical outcome and the CI that mediated the electronic decay. 

The norm of the Dyson orbitals, which provide estimates of the ionization intensities, are relatively independent of the initial geometry (see Figure S6 in the SI). Ensemble properties were therefore computed as unweighted averages of all trajectories. Uncertainties associated with the ensemble size were estimated with the bootstrap method using 1000 resamples. 

The trajectories with exactly three (instead of four) H atoms closer than 2.9\,Å (the sum of the van der Waals radii of H and C) to their nearest C atom are classified as having undergone \ce{H}-loss. Similarly, those with exactly two H atoms closer than 2.9\,Å to their nearest C atom are taken as \ce{H2}- or \ce{2H}-loss trajectories. 

Trajectories with all four H atoms at a distance of less than 2.9\,Å from their nearest C atom, and with three H atoms closer to one C atom than to the other, are considered to be in the ethylidene form. We note that this geometrical criterion differs from that employed in Ref.~\citenum{Joalland-2014}. The structures termed therein the ``bridged minimum" and the ``ethylidene geometry" (where the latter was identified as a transition state) would in the present work both be classified as ethylidene: for both structures, three H atoms are closer to one of the C atoms than to the other. The fractions of trajectories that are in the ethylidene form is defined as the ethylidene isomer population.

To investigate the correlation between the photochemical outcome and the CI that mediated the electronic decay to the cationic ground state, the \ce{D1}/\ce{D0} hopping geometries were clustered with the optimized \ce{D1}/\ce{D0} MECI structures. This was done by computing the RMSDs of the former geometries relative to the latter using the \verb|rmsd| Python package\cite{rmsd}. The pairs of Cartesian coordinates were aligned using the Kabsch algorithm\cite{Kabsch-1976} for removing rotation and translation. Mirroring was also considered in the minimization of the RMSD. The trajectories were assigned according to the smallest RMSD. 

\section{Results and discussion}

This section starts with a description of the ionization/excitation scheme and the electronic population decays. Following this, an overview of the simulated dissociation and isomerization dynamics is provided. The remainder of the section is then concerned with exploring the correlation between the photochemical outcome and the CIs mediating the preceding electronic decay. In pursuance of this, the excitation-energy dependence of the branching between planar and twisted relaxation pathways is reexamined, and optimized \ce{D1}/\ce{D0} MECIs are discussed. 

\subsection{Ionization and excitation, and ensuing electronic relaxation}
\begin{figure}
    \centering
    \includegraphics{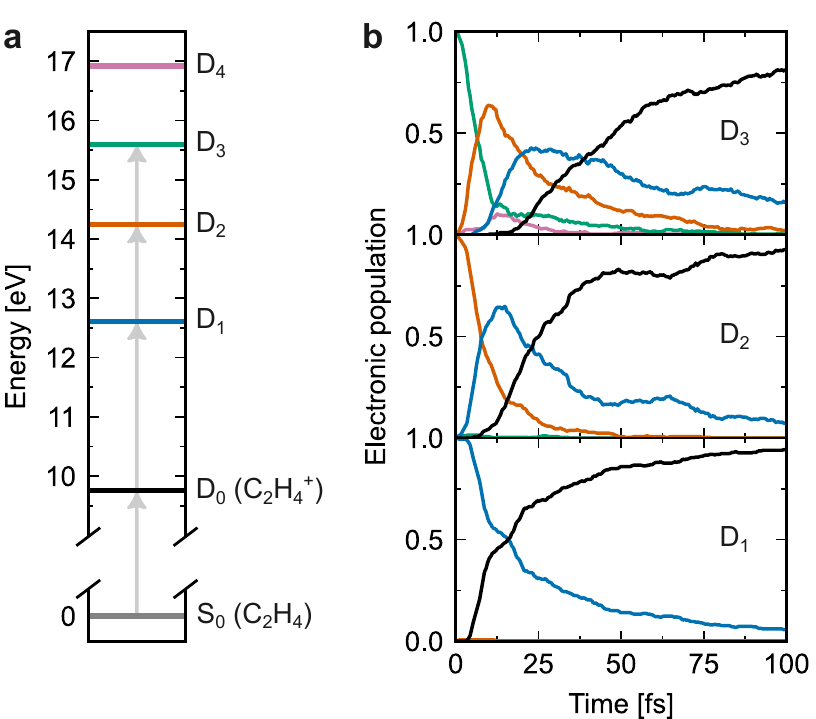}
    \caption{(a) Ionization and excitation of neutral ethylene in its ground electronic state (\ce{S0}) to the four lowest-energy cationic electronic states (\ce{D0}-\ce{D3}). The horizontal lines show the computed energies of \ce{D0}-\ce{D4} at the Franck-Condon point (the optimized geometry of the neutral ground state). (b) Time-dependent adiabatic electronic populations following ionization and electronic excitation to \ce{D3} (top), \ce{D2} (middle) and \ce{D1} (bottom). 
    }
    \label{fig:fig1}
\end{figure}

This work simulates the dynamics upon photoionization and electronic excitation of neutral ethylene to the four lowest-energy cationic states \ce{D0}-\ce{D3} (Figure \ref{fig:fig1}a). The photoelectron spectrum computed for these transitions (Figure S7 in the SI) shows good agreement with a previously reported experimental one\cite{Branton-1970}. The cationic ground state is characterized by a singly occupied $\pi$ orbital, and \ce{D1}-\ce{D3} are all characterized by singly occupied $\sigma$ orbitals. Several recent experimental investigations employed pump-probe schemes in which the pump ionizes and excites neutral ethylene to one\cite{Zinchenko-2021} or several\cite{Tilborg-2009, Ludwig-2016, Vacher-2022, Lucchini-2022} of these electronic states, allowing a comparison between theory and experiment. 

Figure \ref{fig:fig1}b shows the time evolution of the adiabatic electronic state populations. The electronic relaxation is ultrafast: 50\% of the population has decayed to \ce{D0} within 48, 25 and 16\,fs following excitation to \ce{D3}, \ce{D2} and \ce{D1}, respectively. The \ce{D1}$\rightarrow$\ce{D0} electronic relaxation time determined herein is somewhat slower than that reported in the work of Zinchenko et al. \cite{Zinchenko-2021}\cite{Zinchenko-2021}, where an experimental \ce{D1}$\rightarrow$\ce{D0} decay time of 6.8$\pm$0.2\,fs was obtained by attosecond transient absorption spectroscopy at the carbon K-edge. The experimental observable, which is not directly comparable with the adiabatic electronic decay, was simulated by AIMS combined with X-ray spectroscopic calculations. Nevertheless, we consider the presently calculated timescale to be in qualitative agreement with this measurement and to be within the precision range expected for the surface hopping method. Indeed, the electronic population decay can depend on the scheme used to approximate the non-adiabatic couplings\cite{Merritt-2023}. 

\subsection{Dissociation and isomerization: yields and time scales}
\begin{figure}
    \centering
    \includegraphics{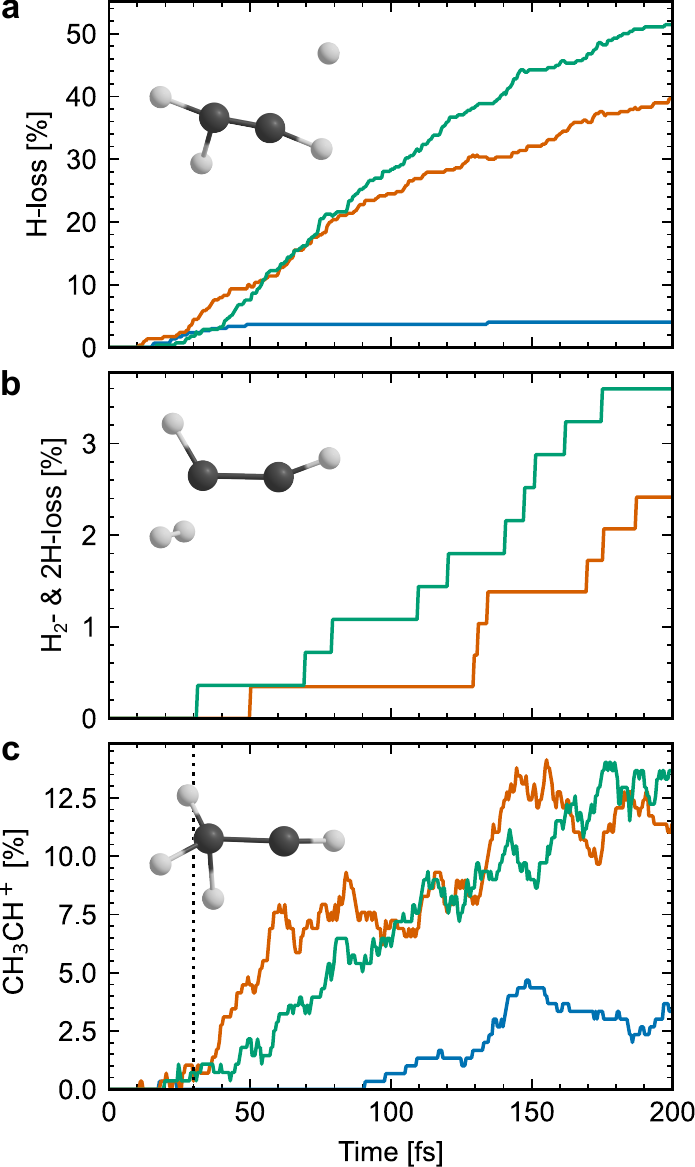}
    \caption{(a) Time-dependent \ce{H}-loss yield following initial population of \ce{D1} (blue), \ce{D2} (orange) and \ce{D3} (green). (b) Same as panel a, but for \ce{H2}- and \ce{2H}-loss. (c) Same as panel a, but for the ethylidene isomer population. The vertical dashed line indicates a previously experimentally inferred ethylene-ethylidene isomerization time from Ref.~\citenum{Ludwig-2016}.}
    \label{fig:fig2}
\end{figure}

In terms of nuclear dynamics, a significant amount of \ce{C-H} bond dissociation and ethylene-ethylidene isomerization is observed in the simulations. These photochemical events take place almost exclusively after the ultrafast electronic relaxation to \ce{D0} (see Table S1 in the SI). The time evolution of the \ce{H}-loss yield, the \ce{H2}-loss yield, and the ethylidene isomer population are shown in Figures \ref{fig:fig2}a, \ref{fig:fig2}b and \ref{fig:fig2}c, respectively. The results are displayed separately for ionization and excitation to \ce{D1}, \ce{D2}, and \ce{D3}. The contributions from excitation to each of \ce{D0}-\ce{D3} to the simulated photoelectron spectrum (Figure S7 in the SI) are largely non-overlapping, making a grouping of the trajectories according to the initial state roughly equivalent with a grouping according to the excitation energy. None of the trajectories initiated on \ce{D0} dissociates or isomerizes, and no \ce{H2}- or \ce{2H}-loss is observed for those initiated on \ce{D1}. In the following, the time scale and mechanism of each photochemical channel are discussed in more detail. 

The time-dependent \ce{H}-loss yields reported in Figure \ref{fig:fig2}a are defined as the percentage of the trajectories that have exactly three \ce{H} atoms attached to a \ce{C} atom at a given time step (see Theoretical Methods for details). At the end of the simulation time (200\,fs), $\sim$40 and $\sim$50\% of the trajectories initiated on \ce{D2} and \ce{D3}, respectively, have undergone \ce{H}-loss. The results are qualitatively different upon \ce{D1} excitation, where the yield saturates at only 4\% after $\sim$50\,fs.

The simulations thus indicate a strong excitation-energy dependence of the \ce{H}-loss yield, in line with experimental data \cite{Lucchini-2022}. They employed an XUV-pump NIR-probe scheme and varied the composition of the initial electronic superposition by selecting single harmonics. Approximate \ce{H}-loss yields can be extracted from a graph in Ref.~\citenum{Lucchini-2022} displaying static (XUV-pump only) \ce{C2H3+} ion yields. These were $\sim$10\% upon excitation with H9, which populates mainly \ce{D0} and \ce{D1}, and $\sim$40\% upon excitation with H11 and H13, which populate \ce{D2} and \ce{D3} more efficiently. Note that the above-cited experimental values are the asymptotic (long-time) yields; the presently modeled \ce{H}-loss yields may increase beyond the simulation time window. 

In addition to the experimental results described above, Lucchini et al. \cite{Lucchini-2022} reported mixed non-adiabatic and adiabatic dynamics simulations with 4, 41 and 55\% \ce{H}-loss upon excitation to \ce{D1}, \ce{D2} and \ce{D3}, respectively\cite{Lucchini-2022}. The first part of the dynamics was described with the surface hopping method with a CASSCF(11e,9o) treatment of the electronic structure. According to the authors, this active space is not suited to describe fragmentation because of the limited number of active orbitals. They, therefore, switched to adiabatic dynamics at the DFT/B3LYP level of theory after relaxation to \ce{D0}. The adiabatic dynamics were propagated for 2\,ps, but the time scale and mechanism of \ce{H}-loss was not described. In the present work, a statistically significant amount of \ce{H}-loss is predicted despite maintaining the same electronic structure method throughout the non-adiabatic dynamics simulations, which we attribute to the larger CASSCF active space.

Corresponding data for the competing \ce{H2}- and \ce{2H}-loss channel is shown in Figure \ref{fig:fig2}b. On the time scale of the simulations, 0, 2.4, and 3.6\% of the population initiated on \ce{D1}, \ce{D2} and \ce{D3}, respectively, undergo \ce{H2}- or \ce{2H}-loss. These yields are approximately one order of magnitude lower than the experimental \ce{C2H2+} ion yields extracted from data reported \cite{Lucchini-2022} ($\sim$20\% for all three harmonics\cite{Lucchini-2022}). To explain this discrepancy, we hypothesize that \ce{H2}- and \ce{2H}-loss continues on a time scale exceeding the simulated 200\,fs. The amount of \ce{H2}-loss from the ethylidene isomer may depend on the cumulative time spent in the latter form, which keeps increasing (Figure \ref{fig:fig2}c). In the mixed non-adiabatic and dynamics simulations of \cite{Lucchini-2022} (discussed above), 1, 16 and 13\% \ce{H2}- and \ce{2H}-loss were reported 2\,ps after excitation to \ce{D1}, \ce{D2} and \ce{D3}, respectively\cite{Lucchini-2022}. However, as was the case for \ce{H}-loss, the time scales and mechanisms of \ce{H2}- or \ce{2H}-loss were not discussed therein. 

Three distinct \ce{H2}- and \ce{2H}-loss mechanisms are represented in the simulations of the present work (Table \ref{tab:H2_mech}): (I) \ce{H2}-loss from the ethylidene isomer, (II) sequential \ce{2H}-loss from ethylene, and (III) \ce{H2}-loss from ethylene. Note here that the branching ratios between them reported in Table \ref{tab:H2_mech} should be interpreted with care because of the limited statistics. 

\begin{table*}[b]
  \begin{center}
    \caption{Mechanisms of \ce{H2}- and \ce{2H}-loss and the branching ratios between them following ionization and excitation to \ce{D2} and \ce{D3}. The numbers of trajectories corresponding to the percentages are indicated in parentheses. 
    }
    \label{tab:H2_mech}
    \begin{tabular}{l l r r}
    \hline
    & \textbf{Mechanism} & \textbf{\ce{D2}} & \textbf{\ce{D3}} \\ 
      \hline 
      \textbf{I} & \ce{CH2CH2+}$\rightarrow$\ce{CH3CH+}$\rightarrow$\ce{CHCH+} + \ce{H2} & 86\% (6/290) & 40\% (4/278) \\
      \textbf{II} & \ce{CH2CH2+}$\rightarrow$ \ce{CH2CH+} + \ce{H} $\rightarrow$ \ce{CHCH+} + \ce{2H} & 0\% (0/290) & 20\% (2/278) \\
     \textbf{III} &  \ce{CH2CH2+} $\rightarrow$ \ce{CCH2+} + \ce{H2} & 14\% (1/290) &  40\% (4/278)\\ \hline
    \end{tabular}
  \end{center}
\end{table*}

\ce{H2}-loss from the ethylidene isomer dominates following ionization and excitation to \ce{D2}. Upon initial population of \ce{D3}, \ce{H2}-loss from ethylidene and ethylene contribute equally, with a lesser contribution from \ce{2H}-loss from ethylene. To our knowledge, only mechanism (I) has been discussed previously in the literature in the context of the ultrafast dynamics of the ethylene cation, and this mechanism has previously only been supported by static calculations. \ce{2H}-loss exhibits a higher threshold from the cationic ground state (7.3\,eV\cite{Stockbauer-1975}) compared to \ce{H2}-loss (2.6\,eV\cite{Stockbauer-1975}), and it is therefore reasonable that, if the former contributes, it does so only upon high-energy excitation. These three mechanisms cannot be distinguished by experimental schemes relying solely on the \ce{C2H2+} fragment as a signature for \ce{H2}- or \ce{2H}-loss.  

We now turn our attention to the ethylidene isomer, which is involved in \ce{H2}-loss mechanism (I). Figure \ref{fig:fig2}c  shows the ethylidene isomer populations, defined as the percentage of the trajectories that are in the \ce{CH3CH+} form (see Theoretical Methods for definition) at a given time step. Ethylene-ethylidene isomerization is reversible on the time scale of the simulations, as indicated by the nonmonotonic increases in the populations in Figure \ref{fig:fig2}c and as inferred by visual inspection of the time evolution of the nuclear geometries along selected trajectories using the molden program\cite{Schaftenaar-2017}. 

The time-dependent \ce{CH3CH+} populations are rather similar for \ce{D2} and \ce{D3} excitation. They start to build up at $\sim$30\,fs, and thereafter continue to increase until the end of the simulation time, where they reach $\sim$10-15\%. The population increase following \ce{D2} excitation is more uneven than that following \ce{D3} excitation. Upon ionization and excitation to \ce{D1}, the modeled onset of ethylene-ethylidene isomerization occurs later, at $\sim$90\,fs, despite the faster electronic relaxation to \ce{D0}. At $\sim$150\,fs, a local maximum, at $\sim$ 5\%, is thereafter seen.

The time scale of ethylene-ethylidene isomerization has previously been inferred from indirect experimental measurements employing XUV-pump NIR-probe schemes\cite{Tilborg-2009, Ludwig-2016}. In these works, the \ce{CH3+} ion --- assumed to be formed by a probe-induced breakup of the \ce{C=C} bond in \ce{CH3CH+} --- was taken as a signature of the isomerization. From a fit to the \ce{CH3+} ion yield, Ludwig et al. \cite{Ludwig-2016} extracted an upper bound of the ethylene-ethylidene isomerization time of 30$\pm$3\,fs\cite{Ludwig-2016}, which is indicated in Figure \ref{fig:fig2}c by the dashed vertical line. They define it as ``the time that the nuclear wave packet needs to reach a region of the potential energy surface where the isomer population probability is maximum"\cite{Ludwig-2016}. As evident from Figure \ref{fig:fig2}c, the present work's modeled isomerization time differs considerably from the above cited experimentally inferred one: 30\,fs after ionization and excitation to \ce{D1}, \ce{D2} and \ce{D3} (the states mainly populated experimentally by the XUV-pump in Ref.~\citenum{Ludwig-2016}), the simulated ethylidene isomer population probabilities are essentially zero. This large discrepancy may indicate an alternative interpretation of the time dependence of the experimental \ce{CH3+} ion yield. 

\subsection{Dissociation and isomerization: role of conical intersections}
After having described the overall yields and time scales of \ce{H}-loss, \ce{H2}- and \ce{2H}-loss, and ethylene-ethylidene isomerization in the preceding subsection, the present subsection aims to explore if, and in that case how, these yields depend on the CIs that mediated the electronic decay to \ce{D0} (recall here that most dissociation and isomerization events take place on \ce{D0}). As a first step, hopping geometries and \ce{D1}/\ce{D0} MECIs are presented. 

\begin{figure}
    \centering
    \includegraphics{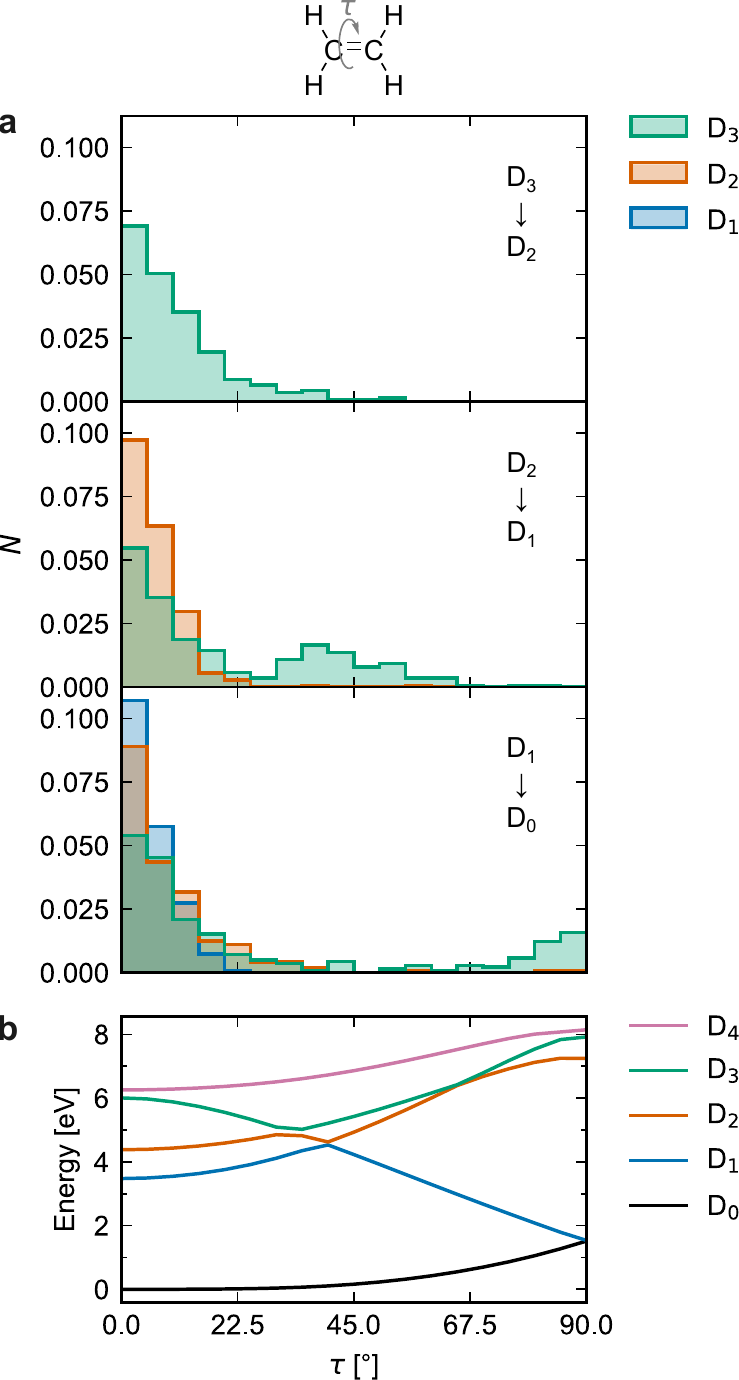}
    \caption{(a) Distribution of dihedral angles at the \ce{D3}$\rightarrow$\ce{D2} (top), \ce{D2}$\rightarrow$\ce{D1} (middle), and \ce{D1}$\rightarrow$\ce{D0} (bottom) hopping geometries following excitation to \ce{D1} (blue), \ce{D2} (orange) and \ce{D3} (green). (b) Potential energy curves along the \ce{H-C-C-H} dihedral coordinate $\tau$ obtained by constrained optimization on \ce{D0}.}
    \label{fig:fig3}
\end{figure}

The non-adiabatic transitions between the low-lying states of \ce{C2H4+} have previously been shown to proceed via two distinct classes of CIs characterized by planar and twisted geometries\cite{Joalland-2014, Ludwig-2016, Lucchini-2022}. MECIs are presented in detail below, after a discussion of hopping geometries. Figure \ref{fig:fig3}a shows the distribution of \ce{H-C-C-H} dihedral angles $\tau$ for the \ce{D3}$\rightarrow$\ce{D2}, \ce{D2}$\rightarrow$\ce{D1}, and \ce{D1}$\rightarrow$\ce{D0} hopping geometries upon initial population of \ce{D1}, \ce{D2} and \ce{D3}. The graphs display the dihedral angle at each trajectory's first hop between the given pair of states (many trajectories undergo back-hopping and thereafter transition between the same pair of states again). Upon excitation to \ce{D3}, the \ce{D3}$\rightarrow$\ce{D2} transitions are centered around planar geometries ($\tau=0^\circ$). The wavepacket then splits, with one part undergoing the \ce{D2}$\rightarrow$\ce{D1} non-adiabatic transitions around planar, and the other around $\sim45^\circ$ twisted, geometries. Further twisting thereafter occurs, and the \ce{D1}$\rightarrow$\ce{D0} transitions take place around planar and 90$^\circ$ twisted geometries. In contrast, upon excitation to \ce{D1} and \ce{D2}, the system is brought to \ce{D0} almost exclusively via planar CIs. 

The torsional relaxation channel thus comes into play essentially only upon excitation to \ce{D3}. This can be rationalized partly from the dihedral scan in Figure \ref{fig:fig3}b, where the $\sim45^\circ$ twisted \ce{D2}/\ce{D1} and the 90$^\circ$ twisted \ce{D1}/\ce{D0} CIs are visible. The latter is a symmetry-induced (Jahn-Teller\cite{Jahn-1937}) intersection. Around $\tau=0^\circ$, where the Franck-Condon (FC) point is located upon vertical ionization and excitation of neutral ethylene, torsional barriers are present on \ce{D2} and \ce{D1}, but not on \ce{D3}. In agreement with our simulations, twisted CIs were involved almost exclusively upon \ce{D3} excitation in previous non-adiabatic dynamics simulations initiated around the equilibrium geometry of the neutral species. \cite{Ludwig-2016, Lucchini-2022}. In Ref. \citenum{Joalland-2014}, where the dynamics were initiated around the optimized geometry of the cation ($|\tau|=18^\circ$), the twisted decay pathway contributed also upon \ce{D1} and \ce{D2} excitations, albeit to a lesser extent than after photoexcitation to \ce{D3}. 

In the following, we focus on the \ce{D1}/\ce{D0} CIs and their effect on the ensuing dissociation and isomerization dynamics. Optimized \ce{D1}/\ce{D0} MECIs and associated population transfers are first described, whereafter MECI-specific dissociation and isomerization yields are presented. 

\begin{figure}
    \centering
    \includegraphics{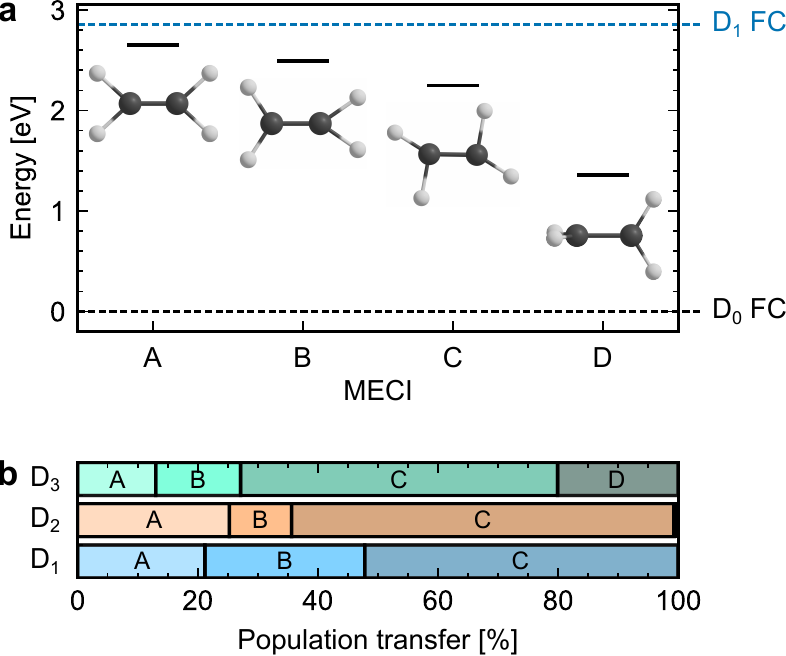}
    \caption{(a) Optimized \ce{D1}/\ce{D0} MECIs. (b) Population transfer for each of the \ce{D1}/\ce{D0} MECIs after excitation to \ce{D3} (top), \ce{D2} (middle) and \ce{D1} (bottom).}
    \label{fig:fig4}
\end{figure}

Four \ce{D1}/\ce{D0} MECIs are identified, of which three are planar (A-C in Figure \ref{fig:fig4}a) and one twisted (D in Figure \ref{fig:fig4}a). All of them are energetically accessible from the \ce{D1} FC point. In terms of molecular geometry, MECI A is accessed by a contraction of the \ce{C=C} bond and a symmetric scissoring of the \ce{H-C-H} angles, and MECI B by asymmetric \ce{H-C-H} scissoring. The two lowest-energy structures are C, located along an \ce{H}-migration coordinate, and D, the 90$^\circ$ twisted CI. MECIs B-D were previously described in Ref. \citenum{Joalland-2014}, and a structure similar to A was identified in Ref. \citenum{Zinchenko-2021}. A geometrical and energetic comparison of the presently found and previously reported MECIs is provided in Figure S8. In addition, CIs along a \ce{C-H} stretch coordinate have been proposed to promote \ce{H}-loss\cite{Sannen-1981, Ludwig-2016}. Whereas we found points along this coordinate that are located on, or close to, the CI seam, our optimizations did not converge on minimum energy structures characterized by one or several elongated \ce{C-H} bonds. 

To obtain the population transfer associated with each MECI, the \ce{D1}/\ce{D0} hopping geometries were clustered with structures A-D as described in the Theoretical Methods section. Upon excitation to \ce{D1}, the hops to \ce{D0} take place close to the MECIs, as indicated by the narrow distribution of RMSDs (see Figure S9 in the SI). The \ce{D1}$\rightarrow$\ce{D0} transitions are less tight following \ce{D2} and \ce{D3} excitation. This, coupled with the geometrical proximity of structures A-C, implies that the non-adiabatic transitions may occur almost equally close to several of these planar MECIs. Moreover, all optimizations initiated from $\sim$150 \ce{D1}/\ce{D0} hopping geometries converged on the lowest energy structures C and D. Although the trajectories clustered with each of the structures within the planar family thus may not be well-separated from each other, this approach does well in separating the trajectories transitioning close to the family of planar MECIs (A-C) from those hopping close to the twisted MECI (D). 

The resulting population transfers for each MECI are displayed in Figure \ref{fig:fig4}b. Regardless of the initially populated state, relaxation through the planar MECI C dominates the decay, with smaller contributions from hops close to structures A and B. These results contrast with those of \cite{Zinchenko-2021}, who discussed the ultrafast \ce{D1}$\rightarrow$\ce{D0} decay upon ionization and excitation of neutral ethylene to \ce{D1} in terms of MECI A only.\cite{Zinchenko-2021} The torsional relaxation channel (MECI D) contributes almost exclusively upon \ce{D3} excitation, in agreement with the data reported in Figure \ref{fig:fig3}a. The population transfers associated with MECIs B-D were previously reported in Ref. \citenum{Joalland-2014}, but a direct comparison cannot be made due to the different initiation of the dynamics.

Next, we turn our attention to the MECI-specific isomerization and dissociation yields and time scales. To rationalize an experimentally observed decrease in the \ce{H2}/\ce{H}-loss ratio at high excitation energies\cite{Kim2007}, Joalland et al. \cite{Joalland-2014} proposed that transition through twisted CIs may hamper ethylene-ethylidene isomerization and thereby disfavor \ce{H2}-loss\cite{Joalland-2014}. The planar MECI C has, conversely, been proposed to promote ethylene-ethylidene isomerization\cite{Joalland-2014, Ludwig-2016}. 

\begin{figure}
    \centering
    \includegraphics{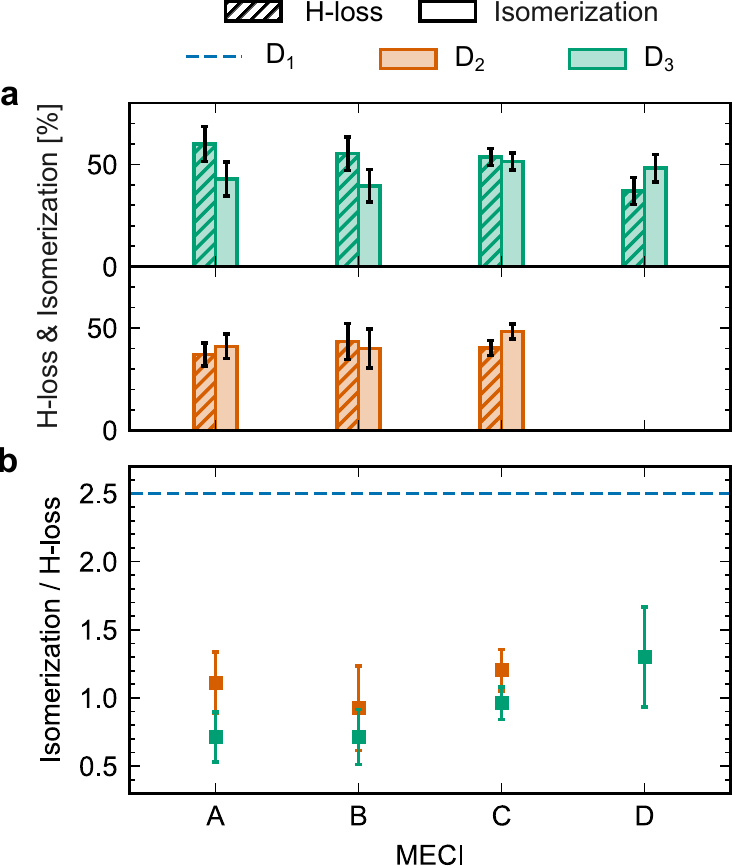}
    \caption{(a) \ce{D1}/\ce{D0} MECI-specific \ce{H}-loss and ethylene-ethylidene isomerization yields following ionization and electronic excitation to \ce{D3} (top) and \ce{D2} (bottom). The error bars show one standard deviation obtained by bootstrapping with 1000 resamples. (b) MECI-specific ethylene-ethylidene isomerization/\ce{H}-loss ratios following \ce{D3} (green) and \ce{D2} (orange) excitation. The blue dashed line shows the overall (across all MECIs) ethylene-ethylidene isomerization/\ce{H}-loss ratio upon excitation to \ce{D1}. The error bars show one standard deviation obtained by bootstrapping with 1000 resamples; no uncertainty is indicated for the ratio upon \ce{D1} excitation because the bootstrapping yielded a highly skewed distribution.}
    \label{fig:fig4.5}
\end{figure}

Figure \ref{fig:fig4.5}a shows the yields of \ce{H}-loss and ethylene-ethylidene isomerization for the trajectories clustered with each of the \ce{D1}/\ce{D0} MECIs. The ethylene-ethylidene isomerization yield is here defined as the percentage of the trajectories that are in the ethylidene form during at least one time step of the dynamics; as noted previously, ethylene-ethylidene isomerization is reversible. Given the small number of \ce{H2}- and \ce{2H}-loss trajectories (only 17 out of 967 trajectories in total), the statistics for this dissociation channel are insufficient for a meaningful comparison of the MECI-specific yields. Similarly, the trajectories initiated on \ce{D1} are excluded from the analysis because of the low \ce{H}-loss yield (12 \ce{H}-loss trajectories out of 299 trajectories for this initial state). Following initial ionization and excitation to \ce{D2}, only two trajectories transition through the twisted MECI D, and these are therefore also excluded from the analysis. 

Following ionization and electronic excitation to \ce{D2}, the ethylene-ethylidene isomerization and \ce{H}-loss yields appear to vary only slightly among the planar MECIs A-C. Figure \ref{fig:fig4.5}a indicates that the variation in the yields associated with the planar MECIs may be larger upon \ce{D3} excitation -- but the statistics do not allow for clear conclusions. 
The similarity of the photochemical outcomes associated with the different planar MECIs A-C may partly be attributed to the geometrical proximity of the latter, combined with the relatively large RMSDs between the hopping geometries and the MECIs. 

\noindent Figure \ref{fig:fig4.5}a suggests that the trajectories relaxing to \ce{D0} at twisted geometries are less inclined to undergo \ce{H}-loss during the 200\,fs simulation time than those transitioning at planar geometries. Ethylene-ethylidene isomerization, in contrast, does not appear to be hampered.  As a result, the ethylene-ethylidene isomerization/\ce{H}-loss ratio (Figure \ref{fig:fig4.5}b) is higher for the trajectories clustered with MECI D than for those clustered with the planar MECIs A-C (although the difference is not statistically significant). 

\begin{figure}
    \centering
    \includegraphics{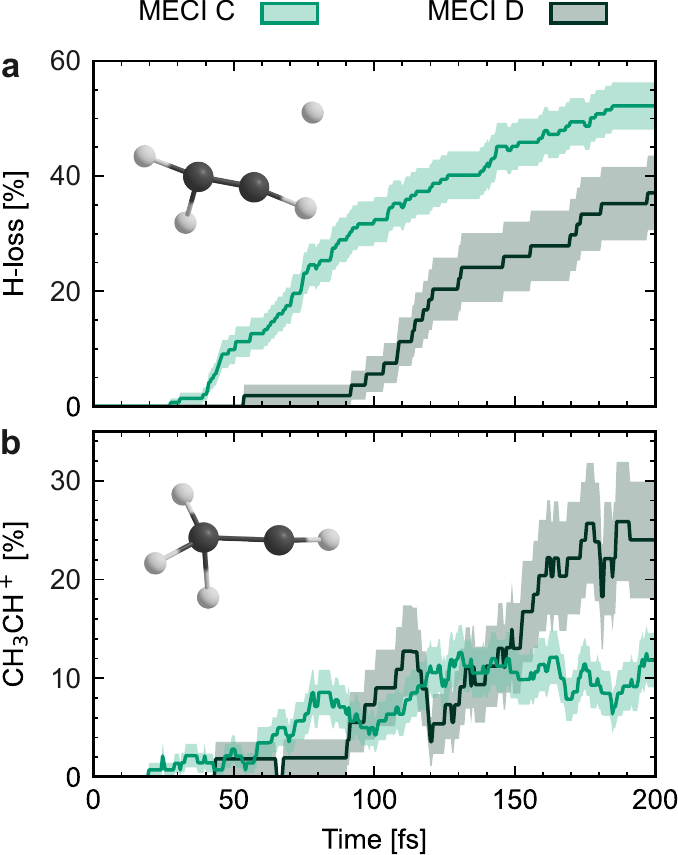}
    \caption{(a) Time-dependent \ce{H}-loss yield for the trajectories transitioning through the planar \ce{D1}/\ce{D0} MECI C (light green) and the 90$^{\circ}$ twisted MECI D (dark green) following excitation to \ce{D3}. The shaded regions show one standard deviation obtained by bootstrapping with 1000 resamples. (b) Same as panel a, but for the ethylidene isomer population.}
    \label{fig:fig5}
\end{figure}

To gain further insight into the dissociation and isomerization dynamics associated with the planar and twisted relaxation pathways, the time evolution of the \ce{H}-loss yields and \ce{CH3CH+} isomer populations are shown for MECI C and D in Figure \ref{fig:fig5}.

\noindent Later onsets of both ethylene-ethylidene isomerization and \ce{H}-loss are observed for the trajectories transitioning at twisted geometries, which may in part be explained by a slower \ce{D1}$\rightarrow$\ce{D0} decay. The slower decay is, in turn, mainly caused by an important amount of back-hopping; almost 90\% of the trajectories that relax to \ce{D0} at twisted geometries hop back to \ce{D1} at least once. This feature has previously been reported in AIMS simulations\cite{Joalland-2014}.  After the delayed onset, the ethylidene isomer population associated with MECI D increases, and at the end of the simulation time, it exceeds that associated with MECI C. The \ce{H}-loss yield, on the other hand, remains lower throughout the simulation time.  

The results presented in Figures \ref{fig:fig4.5} and \ref{fig:fig5} suggest that transitions through the 90$^\circ$ twisted CIs do not hinder ethylene-ethylidene isomerization compared to \ce{H}-loss, which contrasts with the hypothesis of \cite{Joalland-2014}. This indicates that the opening of the torsional relaxation channel may not be the origin of the experimentally observed\cite{Kim2007, Lucchini-2022} decrease in the \ce{H2}/\ce{H}-loss ratio at high excitation energies  --- although we note that the link between ethylene-ethylidene isomerization and \ce{H2}-loss, on which this conclusion is based, is yet to be firmly established. The excitation-energy dependent ethylene-ethylidene isomerization/\ce{H}-loss ratio in Figure \ref{fig:fig4.5}b may be in line with the experimentally observed excitation-energy dependent \ce{H2}/\ce{H}-loss ratios, if (i) the trend persists over time (and with increasing ensemble size), (ii) \ce{H2}-loss following excitation to \ce{D1} occurs, like following \ce{D2} excitation, mainly from the \ce{CH3CH+} isomer, and (iii) the \ce{H2}-loss yield correlates with the \ce{CH3CH+} isomer population.

\subsection{Conclusion}
The present work used extensive surface hopping simulations to explore the mechanisms and time scales of \ce{H}-loss, \ce{H2}- and \ce{2H}-loss, and ethylene-ethylidene isomerization following photoionization and electronic excitation of ethylene. The electronic structure was treated at the CASSCF level with a large active space suited for a description of bond breaking and formation. The simulated \ce{H}-loss yield shows a strong excitation-energy dependence, in agreement with previously reported experimental data\cite{Lucchini-2022}. Further, in line with the mechanism proposed in the literature based on static calculations\cite{Sannen-1981}, the non-adiabatic dynamics simulations reported herein indicate that the \ce{C2H2+} fragment may be formed by \ce{H2}-loss from \ce{CH3CH+}. The simulations indicate further that two additional mechanisms--- sequential \ce{2H}-loss from ethylene and \ce{H2}-loss from ethylene--- may contribute. We note, however, that the statistics for the \ce{H2}- and \ce{2H}-loss channels are limited in the present work, and these results should therefore be interpreted with care. During the simulation time (200\,fs), none of the trajectories initiated on \ce{D1} undergo \ce{H2}- or \ce{2H}-loss, and the \ce{H2}- and \ce{2H}-loss yields following \ce{D2} and \ce{D3} excitation are lower than previously reported experimental asymptotic (long-time) \ce{C2H2+} ion yields\cite{Lucchini-2022}. Possibly, further \ce{H2}- and \ce{2H}-loss occur on a time scale exceeding 200\,fs. 

The ethylene-ethylidene isomerization time predicted by the simulations is considerably longer than those inferred previously from time-dependent \ce{CH3+} ion yields from XUV-pump NIR-probe experiments\cite{Tilborg-2009, Ludwig-2016}. This raises questions about the interpretation of the experimental data. Previous studies relying on this experimental protocol indeed showed that the time-dependent fragment ion yields often are indirect signatures of the XUV-induced dynamics and that complementary dynamics simulations can aid the interpretation\cite{Vacher-2022, Lucchini-2022}. 

The present paper moreover investigated the correlation between the photochemical outcome and the type of CI that mediated the preceding electronic decay (most dissociation and isomerization events take place on \ce{D0}). In contrast to previous hypotheses\cite{Joalland-2014}, the simulations predict that when compared to transitions through the main planar \ce{D1}/\ce{D0} CI located along the \ce{H}-migration coordinate, transitions through the 90$^\circ$ twisted CI do not suppress ethylene-ethylidene isomerization relative to \ce{H}-loss. The results presented herein therefore suggest that the opening of the torsional relaxation channel at high excitation energies may not be responsible for the experimentally observed decrease in the \ce{H2}-loss/\ce{H}-loss ratios\cite{Kim2007, Lucchini-2022}. This conclusion, however, relies on a link between ethylene-ethylidene isomerization and \ce{H2}-loss, which is yet to be firmly established due to the limited statistics for the latter channel in the present work. Possibly in line with the above-cited experimental trend, the simulations indicate that the trajectories initiated on \ce{D1} may have a higher ethylene-ethylidene isomerization/\ce{H}-loss ratio than those initiated on \ce{D2} and \ce{D3}. This effect, however, cannot readily be traced back to different branching ratios between \ce{D1}/\ce{D0} CIs. 

Future simulations could address the long-time dissociation and isomerization dynamics of ionized and electronically excited ethylene, to see if the experimentally observed \ce{H2}- and \ce{2H}-loss yield can be recovered at longer time scales. Such simulations may provide additional insight into the mechanism(s) of \ce{H2}- and \ce{2H}-loss, and reveal whether the higher ethylene-ethylidene isomerization/\ce{H}-loss ratio observed at low excitation energies in the present work translates into a higher \ce{H2}/\ce{H}-loss ratio. 

\section*{Acknowledgements}
The simulations in this work were performed using HPC resources from CCIPL (Le centre de calcul intensif des Pays de la Loire) and from GENCI-IDRIS (Grant 2021-101353). The project is partly funded by the European Union through ERC grant 101040356 (M.V. and T.T.). Views and opinions expressed are however those of the author(s) only and do not necessarily reflect those of the European Union or the European Research Council Executive Agency. Neither the European Union nor the granting authority can be held responsible for them. This work also received financial support under the EUR LUMOMAT project and the Investments for the Future program ANR-18-EURE-0012 (M.V. and L.F.) L.F. acknowledges thesis funding from the Région Pays de la Loire and Nantes University. S.N. acknowledges COST ATTOCHEM for financial support. 
\section*{Supporting Information Available}
Active space; effect of adding dynamic electron correlation; time step convergence; Dyson intensities and photoelectron spectrum; electronic state for dissociation and isomerization; comparison of MECIs; clustering of \ce{D1}/\ce{D0} hopping geometries with MECIs; optimized geometries.

\bibliography{bibliography}

\end{document}